\documentclass[
reprint,
amsmath,amssymb,
aps,
prl,
longbibliography,
floatfix
]{revtex4-2}

\usepackage{graphicx}% Include figure files
\usepackage{color}
\usepackage[english]{babel}
\usepackage[T1]{fontenc}
\usepackage[utf8]{inputenc}
\usepackage{dcolumn}% Align table columns on decimal point
\usepackage{bm}% bold math
\usepackage{hyperref}
\usepackage[capitalise]{cleveref}
\usepackage[mathlines]{lineno}% Enable numbering of text and display math
\usepackage{multirow}
\usepackage{braket}
\usepackage{bigints}
\def\ket#1{| #1 \rangle}

\usepackage{MnSymbol}

\begin{document}

\preprint{APS/123-QED}

\title{Entanglement Transfer in a Composite Electron--Ion--Photon System}

\author{Axel Stenquist$^1$
}
\author{Jakob Nicolai Bruhnke$^1$
}
\author{Felipe Zapata$^2$%\orcidlink{0000-0003-0095-1730}
}
\author{Jan Marcus Dahlström$^{1\,{\text{\dagger}}}$
}

\affiliation{$^1$Department of Physics, Lund University, 22100 Lund, Sweden. \\
$^2$Departamento de Química Física, Universidad Complutense de Madrid, 28040 Madrid, Spain.}

\begin{abstract}
We study how entanglement in photoionization is transferred from an electron-ion pair to an electron-photon pair by fluorescence. Time-resolved von Neumann entropies are used to establish how information is shared between the particles. Multipartite entanglement, between electron, ion and photon, is found on intermediate timescales. Finally, it is shown how a phase-locked two-pulse sequence allows for the application of time symmetry, mediated by strong coupling, to reveal the entanglement transfer process by measuring the photon number and electron kinetic energy in coincidence. 
\end{abstract}
  
\maketitle

\begin{table}[b!]
\begin{flushleft}
  $^\text{\dagger}$ marcus.dahlstrom@fysik.lu.se
\end{flushleft}
\end{table}

\section{Introduction}
Quantum entanglement and decoherence in ultrafast photoionization is a rapidly emerging research field, moving from its foundational ideas 
%foundations
\cite{
pabst_decoherence_2011, 
zhang_photoemission_2014,
pabst_preparing_2016,
you_attosecond_2016,
carlstrom_quantum_2018,
bostrom_time-stretched_2018,
maxwell_entanglement_2022,
vrakking_control_2021},
%atom: electron+ion
%pabst_decoherence_2011, 
%zhang_photoemission_2014,
%pabst_preparing_2016,
%you_attosecond_2016,
%carlstrom_quantum_2018,
%bostrom_time-stretched_2018,
%yu_core-resonant_2018,
%mehmood_coherence_2021,
%laurell_continuous-variable_2022,
%mehmood_ionic_2023,
%ishikawa_control_2023,
%ruberti_bell_2024,
%stenquist_harnessing_2025,
%atom: two electron
%maxwell_entanglement_2022,
% molecule: electron+ion
%vrakking_control_2021,
%vrakking_ion-photoelectron_2022,
%ruberti_quantum_2022,
%nabekawa_analysis_2023,
%berkane_complete_2025
to its first experimental realizations, with
phase-locked attosecond pulse pairs \cite{koll_experimental_2022}, 
laser-assisted photoionization \cite{shobeiry_emission_2024,busto_probing_2022,laurell_measuring_2025}, 
attosecond transient absorption \cite{goulielmakis_real-time_2010}, 
and strong coupling mediated by Free Electron Lasers (FEL) \cite{nandi_generation_2024}.
While the most common objects in studies of entanglement are spins (artificial atoms) and springs (harmonic oscillators) \cite{haroche_exploring_2006}, it is of fundamental interest to also study how entanglement is generated in natural processes. One such example is photoionization of molecules, which has been studied using synchrotron light \cite{akoury_simplest_2007,schoffler_ultrafast_2008}.
In quantum optics, entanglement can be generated between photons and massive particles, {\it e.g.} electron-photon pairs 
\cite{kfir_entanglements_2019,kazakevich_spatial_2024} and  
atom-photon pairs 
\cite{gorlach_quantum-optical_2020,wein_photon-number_2022}.
Ultrashort laser pulses provide a plethora of unique opportunities to study time-dependent entanglement between massive particles, {\it e.g.} electron--electron pairs \cite{bostrom_time-stretched_2018,maxwell_entanglement_2022} and ion-electron pairs generated from atoms 
\cite{pabst_decoherence_2011,zhang_photoemission_2014,pabst_preparing_2016,you_attosecond_2016,carlstrom_quantum_2018,yu_core-resonant_2018,mehmood_coherence_2021,laurell_continuous-variable_2022,mehmood_ionic_2023,ishikawa_control_2023,ruberti_bell_2024,stenquist_harnessing_2025,goulielmakis_real-time_2010,busto_probing_2022,laurell_measuring_2025,jiang_time_2024}
and molecules 
\cite{
akoury_simplest_2007,schoffler_ultrafast_2008,vrakking_control_2021,vrakking_ion-photoelectron_2022,ruberti_quantum_2022,nabekawa_analysis_2023,berkane_complete_2025,koll_experimental_2022,shobeiry_emission_2024}.
For particles with dimensionality $N,M \ge 2$, 
in Hilbert space $\mathcal{H}^{(N)}\otimes\mathcal{H}^{(M)}=\mathcal{H}^{(N\times M)}$, the dimensionality of the entanglement can be quantified by the Schmidt number, $K\le \min(N, M)$ \cite{terhal_schmidt_2000}, 
or by various other measures based on the reduced density matrix, {\it e.g.} the von Neumann entropy of entanglement \cite{haroche_exploring_2006,cruz-rodriguez_quantum_2024}. 
Beyond bipartition, multipartite systems, $\mathcal{H} = \mathcal{H_A}\otimes\mathcal{H_B}\otimes\mathcal{H_C}\otimes...$, have been investigated theoretically \cite{huber_structure_2013} and experimentally for photons \cite{malik_multi-photon_2016}, trapped ions \cite{lanyon_experimental_2014}, and solids \cite{mathew_experimental_2020,fang_amplified_2025}. High-dimensional multipartite systems have received considerable attention \cite{erhard_advances_2020}, and methods for determining the genuine multipartite 
%(or multiparticle) 
entanglement dimension have been proposed \cite{cobucci_detecting_2024}. 
Despite the numerous studies of entanglement in photoionization, including decoherence induced by inter-orbital correlations in atoms \cite{pabst_decoherence_2011} and by nuclear dynamics in molecules \cite{vrakking_control_2021}, 
%\cite{
%atom: electron+ion
%pabst_decoherence_2011, 
%zhang_photoemission_2014,
%pabst_preparing_2016,
%you_attosecond_2016,
%carlstrom_quantum_2018,
%bostrom_time-stretched_2018,
%yu_core-resonant_2018,
%mehmood_coherence_2021,
%laurell_continuous-variable_2022,
%mehmood_ionic_2023,
%ishikawa_control_2023,
%ruberti_bell_2024,
%stenquist_harnessing_2025,
%atom: two electron
%maxwell_entanglement_2022,
% molecule: electron+ion
%akoury_simplest_2007,
%schoffler_ultrafast_2008,
%vrakking_control_2021,
%vrakking_ion-photoelectron_2022,
%ruberti_quantum_2022,
%nabekawa_analysis_2023,
%berkane_complete_2025,
%goulielmakis_real-time_2010,
%koll_experimental_2022,
%shobeiry_emission_2024,
%busto_probing_2022,
%laurell_measuring_2025,
%nandi_generation_2024}, 
two key questions are left unanswered: 
{\it i)} What is the evolution of such entanglement under ideal conditions, and {\it ii)} how can its inevitable transition into the environment be resolved?

%\begin{figure}
%    \centering
%    \includegraphics[width = 0.45\textwidth]{QER_FIG_2_2_2.pdf}
%\caption{\textit{Entanglement transfer process.} Entanglement is transferred by the three-step process schematically illustrated in (a). The corresponding dynamics are presented quantitatively in (b), showing the von Neumann entropy for different partitions of the system, resolved over pulse duration, $\tau \le \tau_\text{max}$, and post-pulse propagation time, $t_f>\tau_\text{max}$. The lines correspond to the entanglement in the: $\alpha\beta$ electron-ion pair, $\alpha\gamma$ the electron-photon pair, $\alpha\beta\gamma$ and $\alpha\beta\gamma\delta$ the multipartite electron-ion-photon system and $\ell\ell'$ the photon modes.} 
%\label{fig1_1}
%\end{figure}

\begin{figure}
    \centering
    \includegraphics[width = 0.45\textwidth]{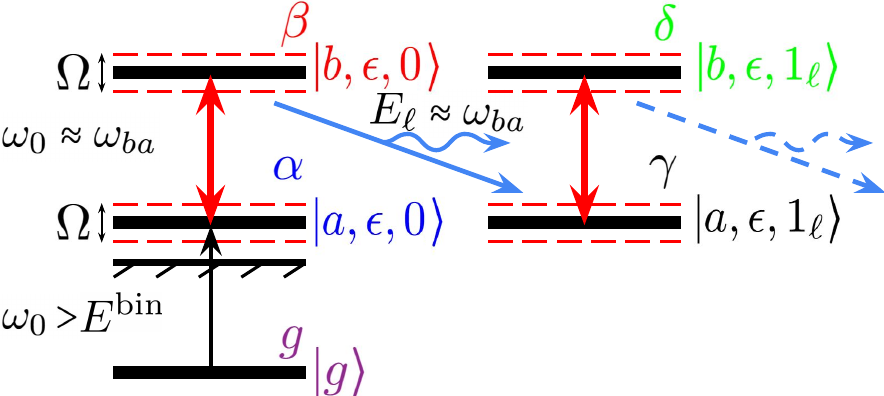}
\caption{
\textit{State coupling schematic.} The atom in its ground state ($g$) is ionized, coupling to the ionic ground state with a free electron ($\alpha$). Subsequently, the field strongly couples the ion, inducing Rabi oscillations to the ionic excited state ($\beta$). The excited state can then spontaneously decay back to the ionic ground state with a fluoresced photon in mode $\ell$ ($\gamma$). Further strong coupling then allows the ion to return to the excited state ($\delta$). Additional spontaneous decay (as denoted by the rightmost arrows) is negligible for the ultrashort laser pulses considered in this work.
} 
\label{FIG_STATES}
\end{figure}

\begin{figure}
    \centering
    \includegraphics[width = 0.45\textwidth]{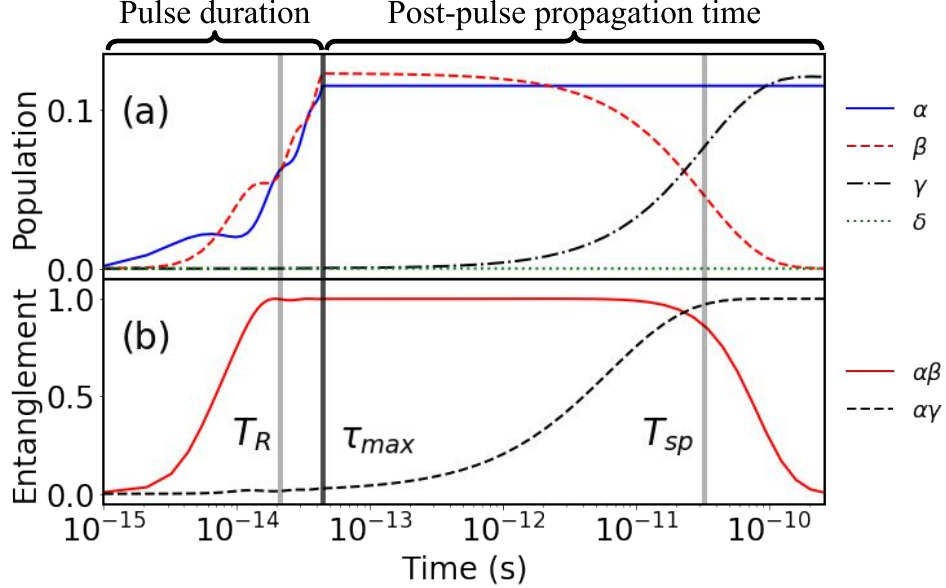}
\caption{
\textit{Time-resolved population and entanglement.} Population (a) and entanglement (b) are resolved over pulse duration $\tau \le \tau_\text{max}$, and post-pulse propagation time, $t_f>\tau_\text{max}$. The population corresponds to the states in \cref{FIG_STATES}. Entanglement is quantified by the von Neumann entropy \cref{eq:vN} between electron and ion, $\alpha\beta$, see \cref{eq:rhoAB}, (red) and between electron and photon number, $\alpha\gamma$, see \cref{eq:rhoAC}, (black dashed). The Rabi period, $T_R$, maximum pulse duration, $\tau_\text{max}$, and spontaneous lifetime, $T_\text{sp}$, are denoted by vertical lines. 
} 
\label{FIG_ENT_POP}
\end{figure}

In this work, we answer these questions by studying the transfer of entanglement from an electron-ion pair to an electron-photon pair.  
While it is known how entanglement is transferred between some prototypical subsystems, {\it e.g.} along spin chains \cite{cubitt_engineering_2008,banchi_optimal_2010,doronin_relay_2018} and by dissipation from smaller systems to larger systems \cite{giddings_quantum_2018}, time-dependent entanglement transfer processes involving two free particles represent a class of much less studied problems. Here, we propose such a physical process with an electron and a photon ejected into two separate continua by photoionization and fluorescence, respectively. 
In order to control the generation of entanglement, we consider photoionization of a helium atom in the strong-coupling regime \cite{nandi_generation_2024}. This implies the usage of an intense laser field with a frequency larger than the binding energy, which is further resonant with an ionic transition:  
\begin{equation}
\mathrm{He}\rightarrow 
\mathrm{He}^+ +\mathrm{e}^- 
\leftrightarrow \mathrm{He}^{+*}+\mathrm{e}^-.
\end{equation}
Thus, absorption of one photon, $\omega_0$, drives the atom from its ground state, $\ket{g}$,  irreversibly into the ionic ground state, $|a\rangle$, creating a free electron with kinetic energy, $E^\mathrm{kin}\approx \omega_0-E^\mathrm{bin}$. According to angular momentum rules, the photoelectron will be a p-wave, $|\epsilon\rangle=|E^\mathrm{kin}p\rangle$. The initial electron--ion state is labelled as $|\alpha(\epsilon)\rangle=|a\rangle\otimes|\epsilon\rangle$. 
After ionization, the ion is further excited because the laser frequency is resonant with an ionic transition, $\omega_0=E_b-E_a$, causing strong coupling at the Rabi frequency, $\Omega(t)$, to the excited ion state $|b\rangle$. We label the excited composite state as $|\beta(\epsilon)\rangle=|b\rangle\otimes |\epsilon\rangle$. 
The entangled electron--ion pair ($\cal{A-B}$) constitutes our primary \textit{bipartite} system $\cal H_A\otimes H_B$ \cite{ruberti_quantum_2022,nandi_generation_2024}, with its corresponding energy levels schematically shown in \cref{FIG_STATES} (left side). Typical ion populations are shown in \cref{FIG_ENT_POP}~(a). 
They imply that $\sim 80\%$ of the atom remains in its neutral ground state. For this reason, it is necessary to ``condition'' on the photoionization event, which means that we will project the total wavefunction on the subspace $\cal H_A\otimes H_B$, and perform a renormalization. In other words, we will only consider the part of the wavefunction containing the formation of an electron-ion pair. This formal procedure is uncontroversial since the coupling from the neutral atom to the electron-ion subspace ceases at the end of the laser pulse.  
%

%While much  attention has been given to ultrafast generation of such electron-ion entanglement, 
%{\it c.f.} Refs.~\cite{pabst_decoherence_2011,zhang_photoemission_2014,pabst_preparing_2016,you_attosecond_2016,carlstrom_quantum_2018,bostrom_time-stretched_2018,yu_core-resonant_2018,mehmood_coherence_2021,laurell_continuous-variable_2022,mehmood_ionic_2023,ishikawa_control_2023,ruberti_bell_2024,stenquist_harnessing_2025,goulielmakis_real-time_2010,busto_probing_2022,laurell_measuring_2025,jiang_time_2024,akoury_simplest_2007,schoffler_ultrafast_2008,vrakking_control_2021,vrakking_ion-photoelectron_2022,ruberti_quantum_2022,nabekawa_analysis_2023,berkane_complete_2025,koll_experimental_2022,shobeiry_emission_2024}, 
However, it must be stressed that electron-ion entanglement corresponds to a fundamentally transient quantum state in nature, since the ion will spontaneously decay by fluorescing a photon given enough time:
\begin{equation}
\mathrm{He^{+*}}+\mathrm{e}^-\rightarrow \mathrm{He^{+}}+\gamma +\mathrm{e}^-.
\end{equation}
Clearly, this implies that the ion will ``decouple'' from the electron and that their entanglement will be lost. A typical decay of the excited ion population ($\beta$) is shown in \cref{FIG_ENT_POP}~(a) occurring on the time scale of $T_\text{sp}\sim 30$ ps.  The resulting electron--photon pair ($\cal{A-C}$) constitutes our second \textit{bipartite} system, $\cal H_A\otimes H_C$, where the ion has fluoresced into a given mode $\ell$:  $\ket{b,\epsilon,0_\ell} \rightarrow \ket{a,\epsilon,1_\ell}$, as schematically illustrated in \cref{FIG_STATES} (right side).   

In our analysis, we will make use of the fact that fluorescence is slow, compared to the time it takes for the electron to travel away from the ion, and also compared to the typical duration of intense laser pulses, $\tau\ll T_\text{sp}$. It is then reasonable to assume that the ion will spontaneously decay under close to ideal (field-free) conditions and that the electron will be unaffected by such decay. This allows us to perform a pure state analysis of the dynamics \cite{mollow_pure-state_1975}. The composite states of ion, electron and photon are labelled 
$\ket{\gamma(\epsilon,\ell)}=|a\rangle\otimes|\epsilon\rangle\otimes|1_\ell\rangle$ 
and 
$\ket{\delta(\epsilon,\ell)}=|b\rangle\otimes|\epsilon\rangle\otimes|1_\ell\rangle$, where $\ell$ labels a continuum of initially empty photon modes, $\ket{0}$.

The goal of this work is to describe the associated time-resolved entanglement-transfer process, presented in \cref{FIG_ENT_POP}~(b), and to quantify its transient dynamics by applying different conditionings and measures. We will also propose an experimental scheme using two phase-locked pulses for detecting and controlling the associated correlations between electrons and photons. We will refer to the composite states by their Greek letters: $\alpha,\beta,\gamma$ and $\delta$, following the schematic energy levels in \cref{FIG_STATES}. 
Atomic units are used, $e=\hbar=m_e=4\pi\epsilon_0=1$, unless otherwise stated.

%\begin{figure}
%    \centering
%    \includegraphics[width = 0.45\textwidth]{QER_FIG_2_2_3.pdf}
%\caption{\textit{State couplings and population dynamics.} (a) Considered states: atomic ground state ($g$), ionic ground and excited states before fluorescence ($\alpha$ and $\beta$), and ionic ground and excited state after fluorescence ($\gamma$ and $\delta$). (b) Population dynamics resolved over pulse duration, $\tau \le \tau_\text{max}$, and post-pulse propagation time, $t_f>\tau_\text{max}$. Lines correspond to the states described in (a).} 
%\label{fig1_2}
%\end{figure}

%\textit{Theory---}
\section{Theory}
Traditional treatments of resonance fluorescence from dressed atoms utilize the Heisenberg picture and the optical Bloch equations to determine the dynamics.
This approach implies that fluorescence causes decoherence of the reduced atom dynamics and that a steady state emerges in the coherent oscillating field \cite{kimble_theory_1976,cohentannoudji_atomphoton_1998}. 
The corresponding Schrödinger picture becomes overwhelming due to the large number of photons emitted during timescales longer than the spontaneous lifetime \cite{mollow_pure-state_1975}. 
In contrast, ultrashort pulses, of much shorter duration than the spontaneous lifetime, $t_1\ll T_\text{sp}$, induce at most one emitted photon per atom. This allows us to conduct a coherent description in the Schrödinger picture. 
The full wave function is expressed as 
\begin{equation}\label{Eq WF}
    \begin{split}
        \ket{\Psi(t)} & \approx g(t)\ket{g} + \int \! d\epsilon \Big\{ \alpha(t,\! \epsilon)\ket{a,\! \epsilon,\! 0}+ \! \beta(t,\!\epsilon)  \ket{b,\!\epsilon,0} 
        \! \\ 
        &+ \! \int \!\! d\ell\left[
     \gamma(t,\!\epsilon,\!\epsilon_\ell)  \ket{a,\!\epsilon,\!1_{\ell}} \!
     +\! \delta(t,\!\epsilon,\!\epsilon_\ell)  \ket{b,\!\epsilon,\!1_\ell} \right] \Big\}, 
    \end{split}
\end{equation}
% up to first order in fluorescence,
where the integrals run over all modes of the field, $\ell$ (including wavevector and polarization), and the photoelectron state, $\epsilon$ (including all quantum numbers). This wave function is then conditioned on the photoionization event: 
$\mathcal{H_A\otimes H_B\otimes H_C}$ to obtain a pure tripartite state of electron $\mathcal{H_A}=\text{span}(\{\ket{\epsilon}\})$, ion $\mathcal{H}_\mathcal{B}^{(2)} = \text{span}(\{\ket{a},\ket{b}\})$, and photon $\mathcal{H_C}=\text{span}(\{\ket{0},\ket{1_\ell}\})$.
Since there is no coupling back to the neutral atom after the laser pulse, $t>t_1$, the states evolve as a pure state in this subspace. 
%Given three subsystems, $\mathcal{H_A}$, $\mathcal{H_B}$ and $\mathcal{H_C}$, entanglement can be moved by \textit{entanglement transfer} from $\mathcal{A}-\mathcal{B}$ to $\mathcal{A}-\mathcal{C}$ (dash denoting entanglement) \cite{cubitt_engineering_2008,banchi_optimal_2010,doronin_relay_2018,giddings_quantum_2018}. \textcolor{red}{, as shown schematically  in Fig.~\ref{FIG_ENT_POP}(a).} 
%The pulse induces transient strong coupling between the ionic ground and excited state $\ket{a}\leftrightarrow\ket{b}$, which generates entanglement between the ionic qubit, 
%$\textcolor{red}{\mathcal{H}_\mathcal{A}^{(2)}=} \textcolor{black}{\mathcal{H}_\mathcal{B}^{(2)} =} \text{span}(\{\ket{a},\ket{b}\})$, and the electron continuum, $\textcolor{red}{\mathcal{H_B}=}\textcolor{black}{\mathcal{H_A}=}\text{span}(\{\ket{\epsilon}\})$,  
%in $\mathcal{H_A\otimes H_B}$ \cite{nandi_generation_2024,stenquist_harnessing_2025}. 
%
%QUQART
%Here, we find it useful to think of a ququart consisting of ion and fluorescence photon number tensor products: $\mathcal{H}^{(4)} = \text{span}(\{\ket{a,0},\,\ket{b,0},\,\ket{a,1},\,\ket{b,1}\})$, which is coupled to the electron and photon mode double continua, $\{\ket{\epsilon},\ket{\ell}\}$.
%
%After the transient multipartite entanglement $\mathcal{A-B-C}$, a secondary bipartite system is reached, and the entanglement reduces to $\mathcal{A-C}$, consisting of our electron coupled to a photon.
%

%Model system: 
\subsection{Model system parameters}
We consider photoionization of helium, $\ket{g}=\ket{1s^2}$, where a coherent Gaussian field is resonant with the ionic 1s-2p transition, $\omega_0 = 40.8$ eV, and has a peak intensity, $I_0 = 1.25 \cdot 10^{13}$ W/cm$^2$, corresponding to the peak Rabi frequency $\Omega_0 = z_{ba}E_0 \approx 0.2$ eV (with corresponding period $T_R \approx 21$\,fs). Given a maximal pulse duration of $\tau_\text{max} = 44$ fs, the corresponding pulse area then extends to $6\pi$. 
%The \textcolor{red}{effective} Rabi period is $T_R\approx 25$ fs. 
Relativistic effects are neglected, which is a good approximation for the ultrashort pulses under consideration \cite{nandi_generation_2024}. In the present work, this non-relativistic assumption implies that the initial {\it singlet} symmetry of the electron spins in the helium ground state, $\ket{\uparrow\downarrow}-\ket{\downarrow\uparrow}$, is conserved during the entire process. Here, our interest lies instead in the entanglement generation in the electron orbital space and its eventual transfer to the photon space. To good approximation, we can neglect both above-threshold ionization and inter-channel correlation effects, which means that the photoelectron remains a pure p-wave during the entire process. 
The coherent driving field $E(t) = \mathcal{E}_c(t) + \mathcal{E}_c^*(t)$ is defined via the positive frequency component $\mathcal{E}_c(t) = E_0 \Lambda(t) \exp(-i \omega_0 t)/2$, with amplitude $E_0$, envelope $\Lambda(t)$ and central frequency $\omega_0$. 
Formally, a unitary transformation is performed to treat the coherent state of the laser field in quantum optics as a $c$-number field, as shown by Mollow in 1975 \cite{mollow_pure-state_1975}. This transformation is exact if the atomic state and coherent state are initially factorizable, which is an excellent assumption for ultrashort laser pulses. The interaction Hamiltonian is then given by $H_I = H_c + H_R$, containing the interaction with the coherent driving field $H_c$, and the quantum interaction term $H_R = i \sum_\ell V_\text{sp}b_\ell^\dagger+\text{h.c.}$, where $V_\text{sp}$ is the spontaneous emission interaction coefficient
% $= z \sqrt{2E_\ell^3/\pi c^3}$ is the angle integrated spontaneous emission interaction coefficient, $E_\ell$ is the fluoresced photon energy 
and $b_\ell$ is the annihilation operator for field mode $\ell$. Further, adiabatic elimination \cite{cohentannoudji_atomphoton_1998} leads to a non-Hermitian Hamiltonian of the type:
\begin{equation} \label{Eq:5lvlH}
\begin{split}
H(t) &=  
\begin{bmatrix}
\frac{-i\Gamma_I(t)}{2} & 0                & 0                                &  0                       &  0                 \\
V_I(t)                  & \epsilon         & \frac{\Omega(t)}{2}              &  0                       &  0                 \\
0                       & \frac{\Omega(t)}{2} & \epsilon \!-\! \frac{i \kappa}{2} &  0                       &  0              \\
0                       & 0                & V_\text{sp}                   & \epsilon\! +\! \epsilon_\ell & \frac{\Omega(t)}{2} \\
0                       & 0                & 0                             & \frac{\Omega(t)}{2}      & \epsilon \!+\! \epsilon_\ell \!-\! \frac{i \kappa}{2} \\
\\
\end{bmatrix},
\end{split}
\end{equation}
where the real part of the diagonal corresponds to the energy (after elimination) and the imaginary part gives the decay rate of the state. The off-diagonals represent the interaction between states. Thus, this  accounts for ionization of the atom by $V_I(t) = z_{ag}E_0\Lambda(t)/2$ (from state index one to two: $g\rightarrow\alpha$) and fluorescence of the ion by $V_\text{sp}$ (from state index three to four: $\beta\rightarrow\gamma$) at the rates $\Gamma_I(t)=2\pi|V_I(t)|^2$ and $\kappa=2\pi|V_\text{sp}|^2$, respectively and for the Rabi coupling by $\Omega(t)/2$ (from state index two to three: $\alpha\leftrightarrow\beta$ and four to five: $\gamma\leftrightarrow\delta$). 
% This allows us to numerically compute the wave function \cref{Eq WF} and validate our analytical result, which we will describe next.
% \textcolor{red}{Next, we will describe the corresponding interaction amplitudes in more detail.}

\subsection{State amplitudes in interaction picture}
The wave function \cref{Eq WF} is computed analytically by applying the propagator $\ket{\Psi(t)} = U(t,t_0)\ket{\Psi(t_0)}$, where $\ket{\Psi(t_0)} = \ket{g}$. The amplitudes are then attained by solving for the terms of the Dyson-like equation, defined as  
\begin{equation}
    U(t,t_0) = U_0(t,t_0) -i \!\int_{t_0}^t \! dt' U(t,t') V(t') U_0(t',t_0),
\end{equation}
where we additionally account for irreversible decay to the continua. 

{\it 0. ---}
The zeroth-order term, $\ket{\Psi^{(0)}(t)} = U_0(t,t_0)\ket{g}$, corresponds to the amplitude of the atomic ground state, which is given by 
\begin{equation}\label{Eq:amp_g}
    \begin{split}
        g(t) =& \exp\left[ -\pi\frac{ \Omega_{ag}^2}{4} \int_{t_0}^{t} dt' \Lambda^2(t') \right],
    \end{split}
\end{equation}
for a flat continuum, following Yu and Madsen \cite{yu_core-resonant_2018}. The procedure is consistent with a slowly-varying envelope approximation applied to the resolvent-operator technique, where field-induced energy shifts are further neglected \cite{stenquist_harnessing_2025}.

{\it 1. ---} The first-order amplitudes
$$
   \ket{\Psi^{(1)}(t)} = -i \!\int_{t_0}^t \! dt' U_R(t,t') H_c(t') U_0(t',t_0) \ket{g},
$$
caused by photoionization to state $\ket{a,\epsilon}$ by the interaction $H_c(t)$ with subsequent ionic Rabi oscillations by the Rabi propagator: 
\begin{equation*}
\begin{split}
U_R(t,t')\ket{a,\epsilon} =& a(t,t')\ket{a,\epsilon}e^{-i(\epsilon_a + \epsilon)(t-t')} \\ +& b(t,t')\ket{b,\epsilon}e^{-i(\epsilon_b + \epsilon)(t-t')},  
\end{split} 
\end{equation*}
which can be computed using the area theorem applied to the ionic two-level system. It yields the Rabi amplitudes $a(t,t') = \cos[\theta(t,t')/2]$ and $b(t,t') = -i\sin[\theta(t,t')/2]$, which depend on the pulse area $\theta(t,t') = \int^t_{t'} dt'' \Omega_{0}\Lambda(t'')$ \cite{allen_optical_1987-1}. The first-order amplitudes, accounting additionally for fluorescence \cite{mollow_pure-state_1975}, read  
\begin{equation}\label{Eq:amp_ab}
    \begin{split}
        \alpha(\epsilon) =& \frac{\Omega_{ag}}{i2} \!\!\int_{t_0}^{t_1}  \!\!dt \; a(t_1,t) \Lambda(t) g(t)
        e^{-\frac{\kappa}{4} t_1 + (i\epsilon+\frac{\kappa}{4}) t} 
        % e^{-\frac{\kappa}{4} (t_1 - t)}
        \\
        \beta(t_f,\epsilon) =& \frac{\Omega_{ag}}{i2} 
        \!\!\int_{t_0}^{t_1}  \!\!dt \;b(t_1,t)\Lambda(t)g(t) 
        e^{-K t_f + (i\epsilon + K) t} ,
        % e^{-K (t_f - t)}
    \end{split}
\end{equation}
where the pulse interacts with the atom during $t\in[t_0,t_1]$. More specifically, we choose $t_1=-t_0=2.5\,\tau$ for a truncated Gaussian pulse shape with $\tau$ being the full-width at half-maximum of intensity. After the end of the pulse, the wave function is further propagated to the final time $t_f \gg t_1$. As already mentioned, the variable $\kappa$ describes exponential decay due to the fluorescence of a photon. Similarly, the variable $K$ gives a spontaneous decay rate, approximated as the dressed ion value $K = \kappa/4$ during the pulse $t\in[t_0,t_1]$, and $K = \kappa/2$ after the pulse $t\in[t_1,t_f]$. Note that $\alpha$ is independent of $t_f$ because the ground state of the ion, $|a\rangle$, can not decay. 

{\it 2. ---} The second-order terms,
\begin{equation*}
\begin{split}
    \ket{\Psi^{(2)}(t)} =&  - \!\int_{t_0}^t \! dt'  \!\int_{t'}^t \! dt'' U_R(t,t'') H_R \\
   & \times U_R(t'',t') H_c(t') U_0(t',t_0)\ket{g},
\end{split}
\end{equation*}
which additionally account for spontaneous emission from the excited ionic state, $|b\rangle$ by quantum interaction $H_R$, followed by continued Rabi oscillations, read 
\begin{equation}\label{Eq:amp_gd}
    \begin{split}
        \gamma(t_f,\epsilon,\epsilon_\ell) &= \frac{i \Omega_{ag}V_\text{sp}}{2}          \int_{t_0}^{t_1}dt 
        \int_{t}^{t_f} dt'  a(t_f, t') b(t', t) 
        \\ 
        &
        \times  \Lambda(t) g(t)  \, e^{\left(i\epsilon_l-K\right)  t'+\left(i\epsilon +\frac{\kappa}{4}\right)t} 
        \\
        \delta(t_f,\epsilon,\epsilon_\ell) &= \frac{i \Omega_{ag}V_\text{sp}}{2}
        \int_{t_0}^{t_1} dt  \int_{t}^{t_1}  dt' b(t_f, t') b(t', t)  
        \\
        &
        \times \Lambda(t)g(t)  \, e^{-\frac{\kappa}{2} t_f+i\epsilon_l t'+\left(i\epsilon +\frac{\kappa}{4}\right)t}.
    \end{split}
\end{equation}
In order to simplify our account of fluorescence, we assume that the photons are ejected radially from a single stationary atom. This allows us to describe the photon modes by a single continuum $\{\ket{\ell}\}$, with
 the angle-integrated spontaneous emission coefficient given by $V_\text{sp} = z_{ba} \sqrt{2E_\ell^{3} / \pi c^{3}}$. Photons with energy $E_\ell$ are emitted with a fluorescence rate $\kappa = 4 z_{ab}^2 E_\ell^3 c^{-3}$. 
 Our account of fluorescence is quantum mechanical in the lowest order, without back action \cite{rzazewski_resonance_1984,florjaczyk_resonance_1985,lewenstein_theory_1986}, assuming that the photons that are ejected will not interfere with the original coherent field. It should be stressed, however, that our approach is rigorous and that such effects could be studied in the future, by performing the inverse unitary transform (displacement) back to the original quantum-optics frame \cite{mollow_pure-state_1975}. Relative energies for photons, $\epsilon_\ell = E_\ell - \omega_0$, and electrons, $\epsilon=E^\text{kin} - \omega_0 + \epsilon_a$, are introduced for more compact notation. Our analytical amplitudes have been successfully verified by numerical propagation of the 5-level non-Hermitian Hamiltonian in Eq.~(\ref{Eq:5lvlH}), and, additionally, a 7-level non-Hermitian Hamiltonian that accounts for higher-order fluorescence processes. 
 %, as shown in the Supplemental Material \cite{Supplemental_Material}.

The time-dependent populations shown in \cref{FIG_ENT_POP}~(a) were computed using the complex amplitudes in \cref{Eq:amp_g,Eq:amp_ab,Eq:amp_gd} as follows: 
$P_g(t)=|g(t)|^2$,
%$P_{i\in\{\alpha,\beta\}}(t) = \int \!\! d\epsilon  \rho_{ii}(t, \epsilon)$
$P_{\alpha}(t) = \int \!\! d\epsilon  \rho_{\alpha}(t, \epsilon)$, using the density $\rho_{\alpha}(t,\epsilon)=|\alpha(t,\epsilon)|^2$, and 
%$P_{i\in\{\gamma,\delta\}}(t) = \int\!\!d\epsilon\!\!\int \!\! d\epsilon_\ell \rho_{ii}(t, \epsilon,\epsilon_\ell)$
$P_{\gamma}(t) = \int\!\!d\epsilon\!\!\int \!\! d\epsilon_\ell \rho_{\gamma}(t, \epsilon,\epsilon_\ell)$, using the density $\rho_{\gamma}(t, \epsilon,\epsilon_\ell)=|\gamma(t,\epsilon,\epsilon_\ell)|^2$, with $P_\beta(t)$ and $P_\delta(t)$ computed analogously.
%with $\rho_{ii}=|i|^2$, and $P_g=|g|^2$. 

\subsection{Time-dependent entanglement measures}
%Measures of entanglement:
The {\it von Neumann entropy} is used to quantify the entanglement between the particles. It is defined as 
%\textcolor{red}{$S_\text{vN}(t) = - \Tr{\tilde \rho(t) \log_2[\tilde \rho(t)]},$}
\begin{equation}
    S_\text{vN}(t) = - \text{Tr}{\tilde \rho(t) \log_2[\tilde \rho(t)]},
    \label{eq:vN}
\end{equation}
%$ \le \log_2(N)$, for $\dim(\tilde\rho)=N$ (of the smaller subspace).
where $\tilde \rho(t)$ is the reduced post-measurement density matrix formed by conditioning and renormalising the full density matrix \cite{haroche_exploring_2006}. It has a maximum value of $\log_2(d)$, where $d$ is the dimension of $\tilde \rho(t)$ and it can be interpreted as the amount of information in the density matrix. For a bipartite pure system, the amount of information is equal for each subsystem, $S_\text{vN}^{\cal (A)}=S_\text{vN}^{\cal (B)}$ \cite{haroche_exploring_2006}. We use this measure to study multipartite entanglement by applying different bipartitions and conditions on the full system. Bipartite association, {\it e.g.} $(\mathcal{H_A} \otimes \mathcal{H_B}) \otimes \mathcal{H_C}$ or $\mathcal{H_A} \otimes ( \mathcal{H_B} \otimes \mathcal{H_C})$, allows us to discuss bipartite entanglement between a particle and a composite system.
% A maximally entangled state has entropy $S_\text{vN}^\text{max} = \log_2(N)$, for $\dim(\tilde\rho)=N$ (of the smaller subspace).
%
For comparison, we also compute the concurrence, another entanglement measure, defined as 
\begin{equation}
    C = \sqrt{2\{1-\text{Tr}\left[\tilde \rho(t)^2\right]\}},
    \label{eq:concurrence}
\end{equation}
which has a maximum value of $\sqrt{2-2/d}$ \cite{haroche_exploring_2006}. Concurrence is monotonically related to linear entropy, $S_\text{lin}=C^2/2$, which is an approximate measure of information in the density matrix.
\subsection{Partitions and conditions}
In order to compute time-dependent entanglements, we will project our pure state on subspaces corresponding to different physical outcomes. These conditioned density matrices are then traced over ``unresolved'' degrees of freedom to form reduced density matrices for a given particle (or composite particles). Time-dependent entanglement measures of this kind are transient quantities in nature, but they are physical in the sense that they can be computed if the corresponding time-dependent reduced density matrix is known. 
\subsubsection{Electron and ion pair ${(\cal A-B})$ }
The von Neumann entropy \cref{eq:vN} between electron and ion is found by first conditioning on photoionization without fluorescence, $\mathcal{H_A\otimes H_B}\otimes\ket{0}$, and by then constructing the reduced ($2 \times 2$) density matrix of the ion 
\begin{equation}
\tilde \rho_{fg}(t) = \frac{1}{N}\int\! d\epsilon f^*g, 
\label{eq:rhoAB}
\end{equation}
where $f,g\in\{\alpha(t,\epsilon),\beta(t, \epsilon)\}$ and $N$ is a normalization factor, such that $\mathrm{Tr}\tilde\rho=1$. In this way the part of the wavefunction that describes fluorescence is rejected, but the decay of the ionic excited state is included by \cref{Eq:amp_ab}, such that the time-dependent electron--ion entanglement will be affected by the spontaneous decay of the ion. 
\subsubsection{Electron and photon pair ${(\cal A-C})$ }
Conversely, the entanglement between electron and photon number is found by conditioning the ground state of the ion, while allowing for fluorescence to any mode, $\mathcal{H_A}\otimes \{\ket{a}\}\otimes \mathcal{H_C}$, using the reduced density matrix, 
\begin{equation}
\tilde \rho_{fg}(t) = \frac{1}{N}\int\! d\epsilon \! \int\! d\epsilon_\ell f^*g,
\label{eq:rhoAC}
\end{equation}
where $f,g\in\{\alpha(\epsilon)F_\delta(\epsilon_\ell),\gamma_\ell(t,\epsilon,\epsilon_\ell)\}$ and $F_\delta(\epsilon_\ell)$ is a delta function. In this way the excited state of the ion is rejected and the ion is fixed at its ground state, while the photon number plays the role of a qubit with photon states taking the values zero or one, $\mathcal{H}^{(2)} = \text{span}(\{\ket{0},\,\ket{1}\})$. In this setup, neither the photon mode nor the electron kinetic energy is detected.

Additionally, there will be dynamics of the information within the resolved photon modes. The entanglement of the photon modes ($\ell\ell'$) with the composite electron and ion system can be studied by conditioning on fluorescence, 
$\mathcal{H_A\otimes H_B\otimes}\{\ket{1_{\ell}}\}$, using the reduced density matrix 
\begin{equation}
\tilde\rho_{\ell,\ell'} = \int\!\! d\epsilon \big[ \gamma(t,\epsilon,\epsilon_\ell)^*\gamma(t,\epsilon,\epsilon_{\ell'}) + \delta(t,\epsilon,\epsilon_\ell)^*\delta(t,\epsilon,\epsilon_{\ell'}) \big],
\label{eq:rholl}
\end{equation}
where the state of the ion has been traced over in addition to the trace over the energy of the electron. 
\subsubsection{Electron, ion and photon system ${(\cal A-B-C})$ }
It is also useful to think of the entropy of a ququart, corresponding to ion and fluorescence photon number tensor products: $\mathcal{H}^{(4)} = \text{span}(\{\ket{a,0},\,\ket{b,0},\,\ket{a,1},\,\ket{b,1}\})$, which is coupled to the electron and photon mode double continua, $\{\ket{\epsilon},\ket{\ell}\}$. The associated entanglement is computed with a density matrix that is defined similarly to \cref{eq:rhoAC}, but with an extended subspace  
\begin{equation}
f,g\in\{\alpha(\epsilon)F_\delta(\epsilon_\ell),\beta(\epsilon)F_\delta(\epsilon_\ell),\gamma_\ell(t,\epsilon,\epsilon_\ell)\}.
\label{eq:fg-ququart}
\end{equation}
for a qutrit, and 
\begin{equation}
f,g\in\{\alpha(\epsilon)F_\delta(\epsilon_\ell),\beta(\epsilon)F_\delta(\epsilon_\ell),\gamma_\ell(t,\epsilon,\epsilon_\ell),\delta_\ell(t,\epsilon,\epsilon_\ell)\},
\label{eq:fg-ququart}
\end{equation}
for a ququart, respectively. In this way, the shared information in the double continua with the discrete dimensions can be studied using a reduced three-by-three or four-by-four density matrix.   

\begin{figure*}
    \centering
    \includegraphics[width=0.9\linewidth]{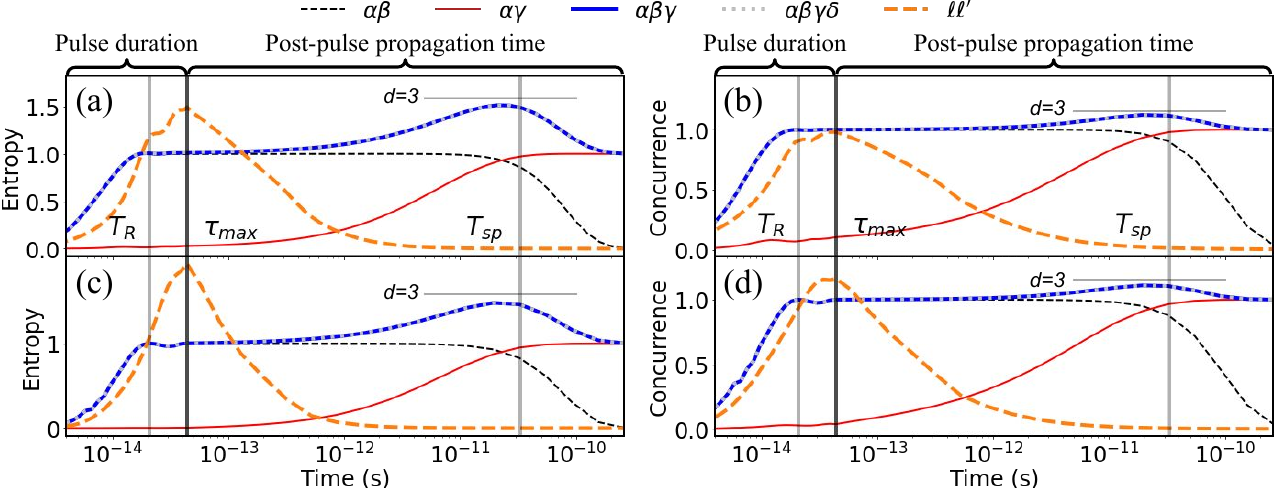}
    \caption{\textit{Entanglement for all partitions, evaluated with entropy of entanglement and concurrence.} Along with the ion-electron entanglement, $\alpha\beta$ (black dashed) and electron-photon entanglement, $\alpha\gamma$, (red), shown in \cref{FIG_ENT_POP}, three additional bipartite entanglements are presented in thicker lines. These are the electron-ion-photon entanglement of the qutrit,  $\alpha\beta\gamma$, and the ququart,  $\alpha\beta\gamma\delta$ states, shown in blue and dotted grey line, and the entanglement of the photon modes (dashed orange). The entanglement is induced by a Gaussian pulse in the top row and a flattop pulse in the bottom row. Entanglement is quantified by the von Neumann entropy $S_\text{vN}$ in the left column and concurrence, $C$, in the right column.} 
    \label{fig:ENT_TOT2}
\end{figure*}

%\textit{Results---} 
\section{Results}

In this section, we present our results for time-dependent entanglement. Of particular interest is how large the transient entanglement can become as the electron and photon move freely away from the ion. For this reason, we will compare two different measures of entanglement with their corresponding maximal values in a given dimension of the reduced space. 
\subsubsection{Electron and ion }
We begin by showing in \cref{FIG_ENT_POP}~(b) how the entanglement between the photoelectron and the ion (labelled $\alpha\beta$) develops in time, computed using \cref{eq:vN} and \cref{eq:rhoAB}. 
%
%As seen in Fig.~\ref{fig1_1}(b), $S_\text{vN}^{(\alpha\beta)}$ rises after approximately one Rabi cycle to the maximal value for the qubit system, $S_\text{vN}^{(\alpha\beta)}\approx 1$. 
%
The time scale is divided into two parts: {\it i)} resolved over pulse duration, up to $\tau \le \tau_\text{max} = 44$ fs, and {\it ii)} post-pulse propagation time beyond the spontaneous lifetime of the ion, $t_f >T_\text{sp}\approx3\times 10^{-11}$ s. In agreement with prior works \cite{nandi_generation_2024,stenquist_harnessing_2025}, we find that the entanglement increases to its maximal value for a qubit system, $S_\text{vN}=1$, in roughly one Rabi period ($T_R$). Beyond this initial rise time, the entanglement remains at its maximal value. This is in contrast to the amount of ionic population, shown in \cref{FIG_ENT_POP}~(a), which keeps increasing further with the pulse duration (until depletion of the atomic ground state, which is not reached at the presented parameter ranges). At longer times, the population of the excited ionic state $(\beta)$ decays due to spontaneous emission. At the same time, we observe how the entanglement between the ion and electron decreases. This decay of the entanglement due to coupling to a larger system is expected \cite{giddings_quantum_2018} and can be qualitatively explained by the decreasing population of the excited ionic state. In more detail, neither the rise of the entanglement at short times, nor the decay of the entanglement at long times, can be explained by the ionic state population alone -- because the entanglement rises earlier and decays later than the excited ionic state. 
% We also notice that the entanglement (label $\alpha\beta$) drops during a shorter time window compared to the decay of the excited state (label $\beta$).     
%
%Strong coupling is induced between states $\alpha$ and $\beta$ during the pulse. %,nandi_generation_2024,stenquist_harnessing_2025,wang_propagation_2023}. 
%After the pulse, the atomic populations remain stationary until the much longer timescale of fluorescence, which leads to population decay in the ion: $\beta\rightarrow \gamma$. 
%Electron and ion:
%
%It then remains stationary beyond the duration of the pulse \cite{nandi_generation_2024,stenquist_harnessing_2025}, until the population in $\beta$ decays, disentangling ion and electron, $S_\text{vN}^{(\alpha\beta)} \rightarrow 0$.  
%
%Electron and photon:
\subsubsection{Electron and photon}
Next, we show in \cref{FIG_ENT_POP}(b) the evolution of the entanglement between the electron and photon number computed using \cref{eq:vN} and \cref{eq:rhoAC} (label $\alpha\gamma$). The entanglement increases to one at the time scale of spontaneous emission (label $T_\text{sp}$). This means that the electron becomes maximally entangled with the qubit photon number system. Compared to the population rise of an ion with a photon in \cref{FIG_ENT_POP}~(a) (label $\gamma$), we notice that the entanglement rises earlier and attains its maximal value before the fluorescence is fully completed.   
 % , $T_\text{sp}\approx 3\times 10^{-11}$ s. %
%Electron, ion, and photon:

\subsubsection{Electron, ion and photon}
Comparing the population transfer in \cref{FIG_ENT_POP}~(a) (labels $\alpha\rightarrow\beta$) with the entanglement transfer in \cref{FIG_ENT_POP}~(b) (labels $\alpha\beta\rightarrow \alpha\gamma$), it is clear that populations are conserved ($\alpha$ and $\beta$ cross at 50\% of their maximal values), while the entanglements are not conserved during the decay of the ion ($\alpha\beta$ and $\alpha\gamma$ cross at $\sim$90\%). 
We therefore expect that during the transfer period, the entanglement in the system will grow beyond that of two coupled qubits, $\mathcal{A-B-C}$. 
This effect can be studied with the ququart entanglement using \cref{eq:vN}, \cref{eq:rhoAC} and \cref{eq:fg-ququart}, shown in \cref{fig:ENT_TOT2}~(a) (label $\alpha\beta\gamma\delta$). This corresponds to a bipartition that is placed such that both the electron energy $\epsilon$ and photon mode $\ell$ are unresolved (traced over), with no further conditioning performed beyond that of photoionisation. We also mention that the qutrit entanglement ($\alpha\beta\gamma$) exhibits identical dynamics because $P_\delta$ is negligible following ultrafast pulse excitation. This implies that the maximal entanglement with this partition can increase beyond $d=2$ value, but must remain below the $d=3$ value for the entanglement. We show that this is true for both the von Neumann entropy \cref{eq:vN} and the concurrence \cref{eq:concurrence} for Gaussian pulses in \cref{fig:ENT_TOT2} panel (a) and (b), respectively. We have verified that similar dynamics occur following excitation by flattop pulses in (c) and (d), respectively. The general behavior of the ququart entanglement (label $\alpha\beta\gamma\delta$) is well reproduced by both measures, with the transient entanglement peak being placed just below the qutrit maximal value (label $d=3$). As expected, the qubit entanglements (labels $\alpha\beta$ and $\alpha\gamma$) are also in qualitative agreement for both measures.    
%
%$S_\text{vN}^{(\alpha\beta\gamma)}$ follows $S_\text{vN}^{(\alpha\beta)}$ at short times, and $S_\text{vN}^{(\alpha\gamma)}$ at long times. 
Thus, we have confirmed that the entanglement between electron and ion is transferred to electron and photon number, but also predicted that the entanglement grows during the ionic decay process to values beyond coupled qubits. 
%
%At intermediate times, $S_\text{vN}^{(\alpha\beta\gamma)}$ grows
%
For the von Neumann entropy, this implies values between $\log_2(2)=1$ and $\log_2(3)\approx1.6$.
%which shows that the entanglement is transiently distributed beyond qubit states. 
%
%Resolved photon modes:
\subsubsection{Resolved photon modes} 
Finally, we turn to the entanglement generated between the photon modes and the rest of the system. We condition on fluorescence, assuming that a photon is detected with a spectrometer (to reveal its mode, $\epsilon_\ell$), while the other parts of the system are unresolved. We compute the entanglement using \cref{eq:rholl} and demonstrate in \cref{fig:ENT_TOT2}~(a) and (c), that the mode entanglement (label $\ell\ell'$) is a delicate quantity that depends on the shape of the laser field. For a Gaussian pulse, we observe that the mode entanglement reaches up to the maximal qutrit entanglement ($d=3$), while for a flattop it exceeds this value. The corresponding concurrence, shown in \cref{fig:ENT_TOT2}~(b) and (d), indicates less entanglement in the modes, with the Gaussian pulse producing a peak at the maximal qubit value, while the flattop pulse has a peak close to the qutrit. Thus, we find that there is good qualitative agreement between the measures and pulses, in the sense that there is a transient peak of the mode entanglement and that this peak is located at the end of the laser pulse. In other words, this is an ultrafast transient entanglement process, which vanishes when the dressing of the ion ceases and the ion decays by spontaneous emission.

\subsubsection{Discussion: Transient entanglement}
\begin{figure}
    \centering
    \includegraphics[width=1\linewidth]{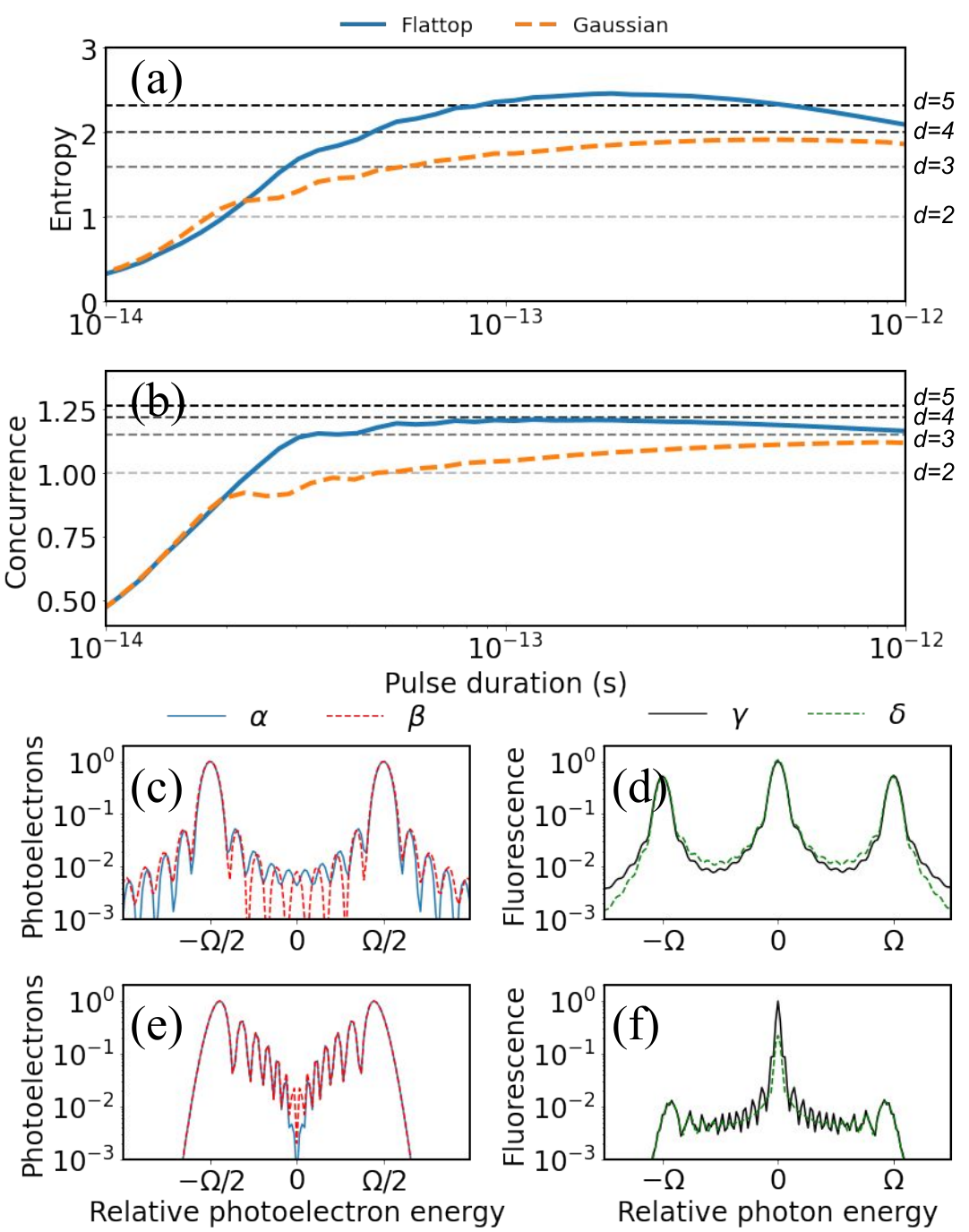}
    \caption{\textit{Entropy of entanglement and concurrence resolved over pulse duration for a flattop and a Gaussian pulse, with photoelectron and fluorescence spectra at maximum entanglement.} Entropy of entanglement is presented in (a) for flattop (blue) and Gaussian (orange dashed) fields, horizontal grey lines denote the maximum entanglement $\log_2(d)$ for dimensions $d=2,3,4,5$. (b) shows the corresponding results for the entanglement measure of concurrence, with maximum value $\sqrt{2-2/d}$. Photoelectron and fluorescence spectra are shown for a flattop pulse in (c) and (d), and a Gaussian pulse in (e) and (f), respectively, for pulse duration $\tau = 200$ fs.
    }
    \label{fig:ENT_ll}
\end{figure}

To further study the mode entanglement, we show in \cref{fig:ENT_ll}~(a) its peak value as a function of pulse duration. Keeping the peak intensity of the pulse constant, while increasing the duration from $\sim$10 fs to $\sim1$ ps, we observe that the mode entanglement has a clear maximum for the flattop pulse on the order of $\sim$100 fs. The fact that there is a maximum is related to the atom being fully ionized, which prevents further generation of entanglement.
% (we have verified that the ion does not get ionized for the pulse durations used in \cref{fig:ENT_ll} \textcolor{green}{Do you mean \cref{fig:ENT_TOT2}?}). 
The maximum is most clearly resolved with the von Neumann entropy (a), where we find that the transient entanglement dimension has increased beyond ququart entanglement ($d=4$), and also beyond qupent ($d=5$). The concurrences show a flatter maximum region for the entanglement, with a lower predicted dimension of the entanglement, close to a ququart ($d=4$). The results for Gaussian pulses show a lower entanglement than the flattop for both entanglement measures. It is observed that the concurrence underestimates the entanglement dimension compared to the von Neumann entropy.

The question then arises as to why flattop pulses can generate a higher amount of transient information in the photon modes compared to the Gaussian pulses. The photoelectron distributions for flattop and Gaussian pulses, in \cref{fig:ENT_ll}(c) and (e), respectively, are quite similar, with two main, equally probable, photoelectron peaks (the so-called ``Autler-Townes doublet'' \cite{autler_stark_1955}, or more specifically the ``Grobe--Eberly doublet'' \cite{grobe_observation_1993}).
We note that large additional fringes are observed between the peaks in the Gaussian case, in agreement with previous investigations \cite{zhang_photoemission_2014}. In contrast, the resonance fluorescence spectra are significantly different for flattop and Gaussian pulses, as shown in \cref{fig:ENT_ll}~(d) and (f), respectively. While the flattop pulse results in three fluorescence peaks with the intensity ratio 1:2:1 (the so-called ``Mollow triplet'' \cite{mollow_power_1969}), the Gaussian pulse yields a more pronounced central peak with two weaker wings \cite{rzazewski_resonance_1984,florjaczyk_resonance_1985,lewenstein_theory_1986}. 
The fact that there is much less probability for emission of photons at frequencies that differ from the central frequency implies that there can be less information encoded in the photon modes for Gaussian pulses.   

% \textcolor{red}{As seen in \cref{fig1_1}(b), $S_\text{vN}^{(\ell\ell')}$ grows during the interaction with the pulse, to a maximum that exceeds that of coupled qubits, $S_\text{vN}^{(\ell\ell')}>1$.
%

In summary, the maximum of the photon-mode entanglement is reached at the end of the pulse (as shown in \cref{fig:ENT_TOT2}) or when the atom is depleted (as shown in \cref{fig:ENT_ll}). It then decreases, as the ion starts to fluoresce at a single frequency, without a resonant driving field. Thus, we find that the information in the photon modes vanishes, while the entanglement between the electron and the photon numbers plays the key role in the transfer process.
%behaves differently from the information in the photon number or electron kinetic energy, which warrants examination of its physical spectra in the asymptotic limit of ionic decay. 
%

%{\it Experimental observables--}   
\subsection{Experimental observables}
\begin{figure*}[t!]
    \centering
    \includegraphics[width = 1\textwidth]{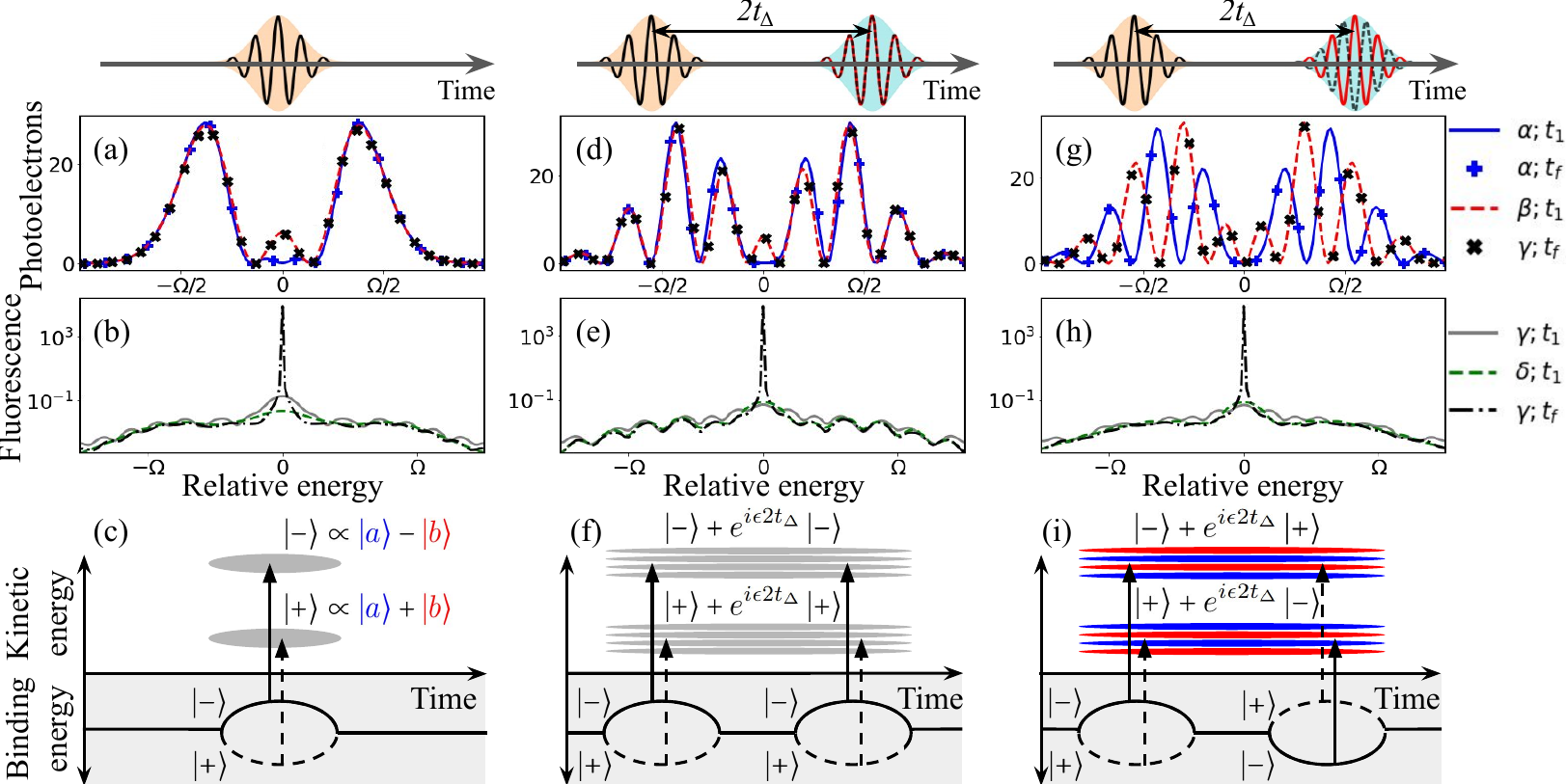}
\caption{\textit{Electron and fluorescence spectra for single Gaussian, even double-Gaussian and odd double-Gaussian fields.} The left column presents results induced by a single Gaussian pulse, with duration $\tau=\tau_\text{max}=44$ fs, (schematically shown at the top). (a) the non-negligible electron spectra at the end of the pulse, $t_1$ ($\alpha(\epsilon)$ blue, $\beta(\epsilon)$ red dashed) and after spontaneous decay, $t_f$ ($\alpha(\epsilon)$ blue +, $\beta(\epsilon)$ black $\times$). (b) The fluorescence spectra of state $\gamma$ and $\delta$ at time $t_1$ (grey and green dashed lines, respectively) and of $\gamma$ after spontaneous decay (dash-dotted black line). A schematic (c) illustrates the formation of the photoelectron peaks in (a) from the dressed states $\ket{-}$ and $\ket{+}$. Corresponding results are shown for an even double-Gaussian field in the middle column, for which the carrier in the second pulse (red line in the top illustration) is in phase with the carrier of the first pulse (black dashed line). The field induces Ramsey-like interference fringes (see (d)), as illustrated in (f). The right column presents the odd double-Gaussian field case where the carriers in the two pulses are out of phase, inducing the avoiding behavior seen in (g), as illustrated in (i).
%
% \textcolor{red}{Electron spectra (top row) and fluorescence spectra (bottom row) at the end of the pulse, $t_1$, and after spontaneous decay, $t_f$, induced by a single Gaussian pulse (a,b), an even double-Gaussian pulse (c,d) and an odd double-Gaussian pulse. Lines correspond to the states shown in \cref{fig1_2}(a).} 
}
\label{fig4}
\end{figure*}
Direct experimental measurement of entanglement in time-dependent quantum systems is a difficult problem. Here, we propose how entanglement transfer can be studied experimentally for our system by considering the correlations between photoelectron spectra with ionic channels and fluorescence photon numbers, and time-resolved fluorescence spectra.   
\subsubsection{Photoelectron spectra}
In the following, we show photoelectron spectra, given by 
% \textcolor{red}{$\rho_{ii}(t,\epsilon)$ for $i\in \{\alpha,\beta\}$ and $\bar \rho_{ii}(t,\epsilon) = \int d\epsilon_\ell \rho_{ii}(t,\epsilon,\epsilon_\ell)$ for $i\in \{\gamma,\delta\}$,}
$\rho_{x}(t,\epsilon)$ for $x\in \{\alpha,\beta\}$ and $\bar \rho_{x}(t,\epsilon) = \int d\epsilon_\ell \rho_{x}(t,\epsilon,\epsilon_\ell)$ for $x\in \{\gamma,\delta\}$, 
%and fluorescence spectra given by $\int d\epsilon \rho_{ii}(t,\epsilon,\epsilon_\ell)$ for $i\in \{\gamma,\delta\}$, 
and how time symmetry of the ultrafast strong coupling can be used to reveal the transfer of entanglement between the particles. 

%Photoelectron spectra: 
The photoelectron spectra induced by a Gaussian pulse are presented in \cref{fig4}(a), where lines show the spectra at the end of the pulse, $t_1$, while markers show times beyond the spontaneous decay, $t_f$.  
The spectra exhibit a doublet, 
% \textcolor{red}{(two peaks separated by the Rabi frequency) in the $\alpha$ (blue) and $\beta$ (red dashed) states at the end of the pulse in agreement with the prediction of Grobe and Eberly in 1993 \cite{grobe_observation_1993}. Weaker internal structures are also observed, due to the smooth pulse envelope \cite{zhang_photoemission_2014}.}
%
generated by ionization from the two dressed states, $\ket{-}$ for $\epsilon>0$ and $\ket{+}$ for $\epsilon<0$, as illustrated in \cref{fig4}(c).
Significant overlap is observed 
$\rho_{\alpha}(t_1, \epsilon)\approx\rho_{\beta}(t_1, \epsilon)$ (except for energies close to $\epsilon=0$), due to both dressed states containing the ground and excited ionic states, $\ket{\pm} \propto \ket{a} \pm \ket{b}$.
States with fluoresced photons are negligible since spontaneous emission is much slower than the ultrafast laser-induced Rabi dynamics. 
After spontaneous decay, $\beta \rightarrow \gamma$, the shape of the electron distribution is conserved $\rho_{\alpha}(t_f,\epsilon)\approx\rho_{\alpha}(t_1,\epsilon)$ (blue +) and $\bar \rho_{\gamma}(t_f,\epsilon)\approx\rho_{\beta}(t_1,\epsilon)$ (black $\times$), while  $\rho_{\beta}(t_f, \epsilon) \rightarrow 0$. 

%Time symmetry: 
Typically, the ion-channel-resolved electron spectra from strong coupling are overlapping, as presented in \cref{fig4}(a), requiring some additional coherent control mechanism to reveal the quantum correlations \cite{nandi_generation_2024}. We recently proposed that time symmetry can be harnessed to generate non-overlapping electron spectra, allowing entanglement to be detected by measuring the electron and the ion state in coincidence \cite{stenquist_harnessing_2025}.  
Here, we consider a field with two delayed and phase-locked Gaussian pulses, centered on $\pm t_\Delta$ with the carrier-envelope phase difference, $\Delta \phi = 2 \omega_0 t_\Delta$ to transfer the entanglement from the internal states of the ion to photon numbers that correlate with the kinetic energy of the electron.

Even field: 
For phase differences, $\Delta\phi = 2n\pi$ ($n$ is an integer), the field is called ``even''. The total pulse area is twice that of a single pulse, since the second pulse is in phase with the first pulse as illustrated at the top of the centre column in \cref{fig4}.
This field yields the electron spectra presented in \cref{fig4}(d). At the end of the pulse, very significant overlap is observed $\rho_{\alpha}(t_1,\epsilon) \approx \rho_{\beta}(t_1,\epsilon)$ (except for $\epsilon\approx 0$). Similarly to the single Gaussian pulse, \cref{fig4}(a), the spectra exhibit a doublet, but this time with additional Ramsey-like interference fringes, yielding several ripples separated by the inverse pulse separation $\pi/t_\Delta$. This is illustrated by the schematic in \cref{fig4}(f), where electrons emitted from the same dressed state interfere as $\sim\ket{-}+\exp(i\epsilon 2 t_\Delta)\ket{-}$, and analogously for $\ket{+}$, inducing the fringes observed in \cref{fig4}(d).
Spontaneous decay transforms $\gamma$ so that $\bar \rho_{\gamma}(t_f,\epsilon) \approx \rho_{\beta}(t_1,\epsilon)$, whereas $\rho_{\alpha}$ remains unchanged. Hence, $\rho_{\alpha}(t_f,\epsilon) \approx \bar \rho_{\gamma}(t_f,\epsilon)$, which means that there is still no evidence for correlations between the electron kinetic energy and the photon number.  

Odd field: 
In contrast, for phase differences, $\Delta\phi = (2n+1)\pi$, the field is called ``odd'', and is illustrated at the top of the right column of \cref{fig4}, where the carriers of the two pulses, shown in black and red, are out of phase.  
%and they have a total area of zero because the second pulse rewinds the state of the atom backward to the initial state. %\cite{stenquist_harnessing_2025}. 
%
This field induces electron distributions, presented in \cref{fig4}(g), which are non-overlapping, both at the end of the pulse $\rho_{\alpha}(t_1,\epsilon) \napprox \rho_{\beta}(t_1,\epsilon)$ and after spontaneous decay $\rho_{\alpha}(t_f,\epsilon) \napprox \bar \rho_{\gamma}(t_f,\epsilon)$. 
The ``avoiding behavior'' of photoelectrons from different ionic states is schematically illustrated in \cref{fig4}(i). Physically, the flip of the sign of the driving laser field reverses the roles of the dressed states, $\ket{-}\leftrightarrow\ket{+}$, such that the eigenvalues correspond to the opposite eigenstate. For this reason, photoelectrons of energy $\epsilon$ will be generated from two different dressed states, $\ket{-}+\exp{(i\epsilon 2 t_\Delta)}\ket{+}$ (for $\epsilon>0$). The ionic state will then oscillate between $\ket{a} \propto \ket{+} + \ket{-}$ and $\ket{b} \propto \ket{+} - \ket{-}$ depending on the electron kinetic energy, $\epsilon$, yielding non-overlapping photoelectron spectra.
Thus, entanglement can be detected by coincidence measurements of the electron kinetic energy and photon number. In other words, we predict that some kinetic energies of the free electron will correlate with the subsequent emission of a photon, while other kinetic energies correlate with no photon being emitted.

\subsubsection{Fluorescence spectra} 
The fluorescence spectra are presented in \cref{fig4}(b), (e), and (h) induced by a single Gaussian, an even double-Gaussian, and an odd double-Gaussian field, respectively. The spectra,  given by $\int d\epsilon \rho_{i}(t,\epsilon,\epsilon_\ell)$ for $i\in \{\gamma,\delta\}$, are similar, having a resonant peak and wings for both $\gamma$ (grey) and $\delta$ (green dashed) at the end of the pulse, but with additional oscillations for $\gamma$. This behavior is in strong contrast with the traditional Mollow triplet structure observed from two-level atoms in resonance fluorescence \cite{mollow_power_1969} (and absorption \cite{stenquist_mollow-like_2024}), but in good agreement with that from smooth pulses  \cite{rzazewski_resonance_1984,florjaczyk_resonance_1985,lewenstein_theory_1986,cavaletto_resonance_2012,moelbjerg_resonance_2012,boos_signatures_2024,vinas_bostrom_time-resolved_2020,liu_dynamic_2024}.  
After the pulse, the excited state decays $\beta\rightarrow\gamma$ by emission of resonant photons with a width corresponding to the spontaneous lifetime (black dash-dotted).
The even double Gaussian case shows additional interference fringes compared to the single Gaussian case.
Finally, we have verified, by numerical simulations of a 7-level non-Hermitian Hamiltonian, that the emission of a second photon yields similar fluorescence spectra. Due to the slow rate of spontaneous emission, such secondary photons have negligible probability and can be safely neglected in our analytical model. We therefore base our experimental scheme on measuring the electron distribution in coincidence with photon number zero or one.

\subsubsection{Discussion: Experimental correlations}
To realize this experimental setup, the phase of the second pulse must be interferometrically stable, so that odd and even fields are distinguishable. The difference in delay between an even and odd pulse is 50 as, making the experiment feasible, as a delay stability of 6 as was recently attained at FERMI \cite{ardini_generation_2024}. The scheme is insensitive to intensity fluctuations as the entanglement is stable after reaching its maximum value, see \cref{fig:ENT_TOT2}. Additionally, we expect that stray photons originating from the laser pulse will dissipate much faster than the slow time scales of fluorescence, limiting the background noise. 

It is noteworthy that a related ``avoiding behavior'' has been reported for electrons in molecular photoionization by two delayed attosecond pulses by Vrakking in 2021 \cite{vrakking_control_2021}. In the molecular case, the ion will undergo vibrational motion with revivals of the initial nuclear wavepacket at well-defined times following ultrafast ionization. Thus, a photoelectron ionized by two pulses with a delay corresponding to the revival time will be connected with the same ionic state superposition, such as $\ket{\epsilon}\otimes(\ket{0}+\ket{1})$ for the ideal two vibrational level case. This results in coherent ionization with a Ramsey-like pattern over the kinetic energy.  
%In the case of an ideal vibrational wavepacket, with the two lowest vibrational levels populated, 
A pulse separation of half the revival time then corresponds to a photoelectron connected with orthogonal ionic states: $\sim\ket{\epsilon}\otimes(\ket{0}+\ket{1})$ and $\sim\ket{\epsilon}\otimes(\ket{0}-\ket{1})$, which can not display any interference over kinetic energy. 
The use of attosecond pulses to create coherence (destroy entanglement) in photoionization is one of the foundational ideas in the field, first proposed by Pabst {\it et al.} in 2011 \cite{pabst_decoherence_2011}. 
%If the ion state is resolved, then the photoelectron will correlate with either $\ket{0}$ or $\ket{1}$ depending on the relative phase $\varphi$ between the two photoemission events, $|1\pm e^{i\varphi}|^2$, for $\ket{0}$ and $\ket{1}$, respectively. 

Coming back to the strong-coupling problem, we conclude that there is a fundamental difference between the physics at play. In our strong-coupling problem, the photoelelectron and ion are always entangled, independently of the pulse separation. In contrast, the amount of entanglement in Vrakking’s setup depends on the relative delay between the pulses. This quantitative difference comes from the pulses being longer than the Rabi period, and shorter than the vibrational period, respectively. Thus, the Ramsey-like fringes in \cref{fig4} (d) in our work do not imply that the photoelectron and ion are coherent, but rather it is a signature of coherence between the laser pulse pair. The avoiding behavior in \cref{fig4} (g) reveals that the photoelectron and ion are indeed entangled.

%
%Complementary flattop and second-photon emission results are found in the Supplemental Material \cite{Supplemental_Material}. 
%

%\textit{Conclusions---}   
\section{Conclusions}
We have studied the time-resolved dynamics of entanglement transfer from photoionization (electron-ion pair) to spontaneous emission (electron-photon pair). 
In order to quantify the transfer process, we have performed different conditions and computed time-dependent von Neumann entropies. 
%
%The fundamental reaction reads
%\begin{equation}\label{Eq:seperability}
%\begin{split}
%    \ket{\Psi (t)}
%    &\rightarrow%{t\, = \, t_1}
%    \Big(\ket{a,\epsilon^{(a)}} + \ket{b,\epsilon^{(b)}}\Big) \otimes \ket{0} %\mathcal{A-B} 
%    \\
%    % \ket{\Psi(t_1)} = 
%    % \big(\ket{a,\epsilon_\alpha} + \ket{b,\epsilon_\beta}\big) \otimes \ket{0_\ell} 
%%    &\xrightarrow{t\, > \, t_1}  
%%    \ket{a,\epsilon^{(a)},0} + \ket{b,\epsilon^{(b)},0} + \ket{a,\epsilon^{(a)},1}  %\mathcal{A-B-C} 
%%    \\
%    % + \ket{b,\epsilon_\beta,1_\ell} \\
%    &\rightarrow%{t\rightarrow \infty}  
%    % \ket{a,\epsilon_a} + \ket{a,\epsilon_b, 1_\ell} = 
%    \ket{a} \otimes \Big(\ket{\epsilon^{(a)}, 0} + \ket{\epsilon^{(b)}, 1}\Big) %\mathcal{A-C} 
%    ,
%\end{split}
%\end{equation}
%where the different separability of the initial and final composite wave function is clear. 
%The electron wave packets, $\ket{\epsilon^{(a)}}$ and $\ket{\epsilon^{(b)}}$, are unchanged by spontaneous emission as $\ket{b,0}\rightarrow \ket{a,1}$. 
%
At the end of the pulse, 
%$t=t_1$, 
the ion and electron are entangled, but the photon number is separable. %$\mathcal{A-B}$. 
As time increases, %$t>t_1$,
we found that all particles (electron, ion, and photon) form a multipartite entangled state. 
%, $\mathcal{A-B-C}$. 
However, after spontaneous decay,
%$t\rightarrow\infty$, 
the wave function is biseparable as the ion becomes factorizable. 
%, $\mathcal{A-C}$.  
%Thus, entanglement is transferred from the primary bipartite ion-electron system, through the tripartite ion-electron-photon system, and finally, to the secondary bipartite electron-photon system, 
%as was demonstrated in \cref{fig1_1}.
%
We considered strong coupling as the mediator of the initial entanglement, which provides full entanglement between the electron and the ion (qubit). The transfer mechanism is of a general nature. 
Additionally, we propose a single-atom experimental scheme based on two pulses to detect the entanglement transfer by coincidence measurements of photoelectron energy and photon number. Such an experiment is made feasible by only requiring the coincidence detection of the photoelectron spectra with photon numbers zero or one.
While only pulse pairs which are in or out of phase are considered, this can be generalised to any phase difference, enabling a broader range of experiments, where the dressed states can be controlled, similar to how different measurement bases are used in Bell inequality experiments.
Our work adds to the field of entanglement transfer, motivating single-atom experiments \cite{haroche_exploring_2006}, driven by ultrashort pulses, to study decoherence and entanglement dynamics between non-identical particles. Our most fundamental observation is that entanglement generated by photoionization is not stationary, but will evolve in time, due to interaction with the environment, {\it e.g.} by fluorescence. 
While some aspects of entanglement are transient in nature, such as the information in the photon modes, others are robust, such as the information in the photon numbers.   
Thus, our work paves the way for novel experiments, where entanglement can be transferred from internal degrees of freedom to the macroscopic world. 

% \begin{equation}
% \begin{split}
%     \ket{-} \propto \textcolor{black}{\ket{a}} - \textcolor{red}{\ket{b}}  \\
%     \ket{+} \propto \textcolor{black}{\ket{a}} +\textcolor{red}{\ket{b}}  \\
%     \ket{-} + e^{i\epsilon 2 t_\Delta} \ket{-} \\
%     \ket{+} + e^{i\epsilon 2 t_\Delta} \ket{+} \\
%     \ket{-} + e^{i\epsilon 2 t_\Delta} \ket{+} \\
%     \ket{+} + e^{i\epsilon 2 t_\Delta} \ket{-} \\
% \end{split}
% \end{equation}

\section{Acknowledgements} 
JMD acknowledges support from the Olle Engkvist Foundation: 194-0734 and the Knut and Alice Wallenberg Foundation: 2019.0154 and 2024.0212, and the Swedish Research Council Grant No. 2024-04247.

\bibliography{QER}

\end{document}